\documentstyle[twoside,fleqn,espcrc2,epsf]{article}

\newcommand{\be}{\begin{equation}}
\newcommand{\ee}{\end{equation}}
\newcommand{\ba}{\begin{array}}
\newcommand{\ea}{\end{array}}
\newcommand{\bea}{\begin{eqnarray}}
\newcommand{\eea}{\end{eqnarray}}
\newcommand{\eqn}[1]{(\ref{#1})}

\def\neb{\hbox{$\overline{\nu}_e\!$ }}
\def\hh{\hat{h}}
\def\hH{\widehat{H}}
\def\hrho{\hat{\rho}}
\def\hp{\hat{\rm p}}
\def\hdmj{\Delta \widehat{M}_j}
\def\hgi{\widehat{\Gamma}_i}
\def\hgj{\widehat{\Gamma}_j}
\def\hmu{\widehat{M}_u}

\def\etal{{\it et al.}}
\def\aa{Astron. Astrophys. }
\def\aj{Astrophys. J. }
\def\ajs{Astrophys. J. Suppl. }
\def\nat{Nature }
\def\np{Nucl. Phys. }
\def\prl{Phys. Rev. Lett. }
\def\pr{Phys. Rev. }

\title{\hfill $\mbox{\small{
$\stackrel{\rm\textstyle DSF-28/99\quad}
{\rm\textstyle astro-ph/9907421\quad}$}}$ \\[.5truecm]
New results in primordial nucleosynthesis}

\author{S. Esposito\address{Dipartimento di Scienze Fisiche and INFN,
Sezione di Napoli, \\ 
Pad. 20 Mostra d'Oltremare, I-80125, Naples, Italy}
\thanks{Talk given by G. Mangano at the {\it 6th San Miniato Topical
Seminar on `Neutrino and Astroparticle Physics'}, San Miniato 1999.}, 
G. Mangano$^{\rm a}$, G. Miele$^{\rm a}$, and O. Pisanti$^{\rm a}$}
       
\begin{document}

\begin{abstract}
We report the results of a new accurate evaluation of light nuclei yields
in primordial nucleosynthesis. The relic densities of $^4He$, $D$ and
$^7Li$ have been numerically obtained {\it via} a new updated version of
the standard BBN code. 
\end{abstract}

\maketitle

\section{Introduction}

Big Bang Nucleosynthesis (BBN) is one of the most powerful tools to study
fundamental interactions since light nuclei abundances crucially depend on
many elementary particle properties, like the number of {\it effective}
neutrino degrees of freedom, $N_\nu$. At the moment, however, recent data
on $^{4}He$ mass fraction, $Y_4$, and Deuterium ($D$) and $^7Li$
abundances, $Y_2\equiv D/H$ and $Y_7\equiv ^7Li/H$, produced during BBN are
controversial, since there are different sets of results, two of them
mutually incompatible: $Y_4^{(l)} = 0.234\pm 0.002\pm 0.005$ and
$Y_4^{(h)} = 0.243\pm 0.003$ \cite{he}, $Y_2^{(l)} = (3.4\pm 0.3) 10^{-5}$
and $Y_2^{(h)} = (1.9\pm 0.4) 10^{-4}$ \cite{deu}, $Y_7^{(l)} = (1.6\pm
0.36) 10^{-10}$ and $Y_7^{(h)} = (1.73\pm 0.21) 10^{-10}$ \cite{li} (for a
brief summary of the experimental situation on primordial abundances see
Ref. \cite{sarkar}).

In spite of this discrepancies, which could be of systematic origin, the
above results for $^4He$ data indicate that one is reaching a precision of
the order of percent, requiring a similar level for the uncertainties on
the theoretical predictions. Besides the corrections to the proton/neutron
conversion rates, which fix at the freeze out temperature $\sim 1~MeV$ the
neutron to proton density ratio, the other main source of theoretical
uncertainty comes from nuclear rates relevant for nuclei formation. In some
cases, these rates are known to well describe the data in a temperature
interval which only partially overlaps the one relevant for BBN, $0.01~ MeV
\leq T \leq 10 ~MeV$. Recent studies \cite{nucin}, however, show that, in
particular for the $^4He$ mass fraction, the effect is at most as large as
the one due to the uncertainty on neutron lifetime $\tau_n$, and smaller
than $1\%$. Therefore it is theoretically justified to look for all sources
of theoretical uncertainty up to this level of precision.

In a previous paper \cite{nprates} we performed a thoroughly analysis of
all corrections (electromagnetic radiative corrections, finite nucleon mass
corrections, plasma effects) to the rates of the processes converting $n
\leftrightarrow p$, i.e. $\nu_e~ n \leftrightarrow e^-~ p$, $\neb~ p
\leftrightarrow e^+~ n$ and $n \leftrightarrow e^-~\neb~p$. Here, we
report on a following work \cite{code}, where we included the above
mentioned corrections in a new updated version of the standard BBN code
\cite{stcode}. This new code was used for integrating the set of equation
of BBN and obtaining the values of the primordial light nuclei yields. From
the comparison of these predictions with the experimental abundances it is
possible to get informations on the effective number of neutrinos and the
final baryon to photon density ratio, $\eta$.

\section{Corrections to Born rates}

As is well known, the key parameter in determining the primordial $^4He$
mass fraction is the value of the neutron to proton density ratio at the
freeze-out temperature $T\sim 1~MeV$, since almost all residual neutrons
are captured in $^4He$ nuclei due to its large binding energy per nucleon.
We shortly summarize the kind of corrections which are studied in detail in
\cite{nprates}. 

The Born rates, obtained in the tree level $V-A$ limit and with infinite
nucleon mass have to be corrected to take into account basically three
classes of relevant effects: 

\begin{itemize} 
\item[i)] order $\alpha$ radiative corrections. These effects have been
extensively studied in literature and can be classified in {\it outer}
factors, involving the nucleon as a whole, and {\it inner } ones, which
instead depend on the details of nucleon internal structure. Actually,
other small effects are expected at higher order in $\alpha$, since the
theoretical value of the neutron lifetime is compatible with the
experimental one, $\tau_n^{ex} = 886.7\pm 1.9~ s$ \cite{pdg}, at 4-$\sigma$ 
level only. These additional contributions are usually taken into account
by eliminating the coupling in front of the reaction rates in favour of
$\tau_n^{ex}$. 

\item[ii)] All Born amplitudes should also be corrected for nucleon finite
mass effects. They affect both the weak amplitudes, which should now
include the contribution of nucleon weak magnetism, and the allowed phase
space. Initial nucleons with finite mass will also have a thermal
distribution in the comoving frame, producing a third kind of finite mass
correction.

\item[iii)] Since all reactions take place in a thermal bath of electron,
positron, neutrinos, antineutrinos and photons, thermal-radiative
corrections should be also included, which account for the electromagnetic
interactions of the in/out particles with the surrounding medium. They can
be evaluated in the real time formalism for finite temperature field
theory. 
\end{itemize}

After solving the BBN set of equations, one can see that for all nuclides
the pure radiative correction provides the dominant contribution, while
the finite nucleon mass effects and the thermal-radiative ones almost
cancel each other.

The total proton/neutron conversion rates were fitted, in the range $0.01~
MeV \leq T \leq 10~ MeV$, to the following functional forms, 
\bea
&&\!\!\!\!\!\!\!\!\!\!\!\! \omega_{n \rightarrow p} (z) =
\frac{1}{\tau_n^{ex}} \exp \left( -q_{np}~ z \right)~ \sum_{l=0}^{13} a_l~
z^{-l}, \label{e:omnp} \\
&&\!\!\!\!\!\!\!\!\!\!\!\! \omega_{p \rightarrow n} (z) =
\frac{1}{\tau_n^{ex}} \exp \left( -q_{pn}~ z \right)~ \sum_{l=1}^{13} b_l~
z^{-l}, \label{e:ompn}
\eea
where $z$ is the dimensionless inverse photon temperature, $z\equiv m_e/T$
and the values of the parameters can be found in \cite{code}. Note that
Eq.~\eqn{e:ompn} is valid only in the range $0.1~ MeV \leq T \leq 10~ MeV$,
because $\omega_{p \rightarrow n} \sim 0$ for $T < 0.1~ MeV$. The fits have
been obtained requiring that the fitting functions differ by less than
$0.1\%$ from the numerical values in the considered range.

\begin{table*}[hbt]
\setlength{\tabcolsep}{1.5pc}
\newlength{\digitwidth} \settowidth{\digitwidth}{\rm 0}
\catcode`?=\active \def?{\kern\digitwidth}
\caption{The predictions on light element abundances obtained with the
numerical code for $\eta = 5\cdot 10^{-10}$}
\label{tab:abund}
\begin{tabular*}{\textwidth}{@{}l@{\extracolsep{\fill}}rrrr}
\hline
                 & $Y_2$
                 & $Y_3$
                 & $Y_4$
                 & $Y_7$ \\
\hline
$\omega_{Tot}$   & $0.3638{\cdot} 10^{-4}$ & $0.1175{\cdot} 10^{-4}$ &
$0.2446$ & $0.2814{\cdot} 10^{-9}$ \\
$\omega_{B}$  & $0.3727{\cdot} 10^{-4}$ & $0.1184{\cdot} 10^{-4}$ &
$0.2550$ & $0.2873{\cdot} 10^{-9}$ \\ 
\hline
\end{tabular*}
\end{table*}

\section{Numerical code}
\subsection{The Equations of BBN}

Denoting with $R$ the universe scale factor, $n_B$ the baryonic density,
$\phi_e \equiv \mu_e/T$ the electron chemical potential, and $X_i$ the
nuclide number densities, $X_i=n_i/n_B$, the BBN set of equations is a
system of coupled differential equations in the previous unknown functions
of time. By expanding the equations with respect to $\phi_e$ and changing
the evolution variable to $z$, after a little algebra one is left with the
following $N_{nuc} + 1$ equations, 
\bea
&&\!\!\!\!\!\!\!\!\!\!\!\! \frac{d \hh}{dz} = \left[ 1-\hH(z,\hh,X_j)~
G(z,\hh,X_j) \right] \frac{3 \hh}{z}, \label{eq:bbn1} \\ 
&&\!\!\!\!\!\!\!\!\!\!\!\! \frac{dX_i}{dz} = G(z,\hh,X_j)~
\frac{\hgi}{z}, \label{eq:bbn2} 
\eea
where we have introduced the dimensionless baryon density, $\hh \equiv
n_B/T^3$, Hubble parameter, $\hH \equiv H/m_e$, and nuclear rates $\hgi
\equiv \Gamma_i/m_e$. The function $G$ in Eq.s~\eqn{eq:bbn1} and
\eqn{eq:bbn2} is
\bea
&&\!\!\!\!\!\!\!\!\!\!\!\! G(z,\hh,X_j) = \left[ \sum_\alpha (4
\hrho_\alpha - z \frac{\partial \hrho_\alpha}{\partial z}) + 4
\Theta(z_D-z) \right. \nonumber \\ 
&&\!\!\! \left. \times \hrho_\nu + \frac 32 ~ \hh \sum_j X_j \right]
\left\{ 3 \left[ \sum_\alpha (\hrho_\alpha + \hp_\alpha) \right. \right.
\nonumber \\
&&\!\!\! \left. \left. + \frac 43 ~ \Theta(z_D-z) \hrho_\nu + \hh \sum_j
X_j \right] \hH + \hh \right. \nonumber \\ 
&&\!\!\! \left. \times \sum_j \left( z \hdmj + \frac 32 \right) \hgj
\right\}^{-1}. 
\eea
In the previous equation $z_D=m_e\, (MeV)/2.3$ is the inverse neutrino
decoupling temperature, $\alpha=e,\gamma$, $\hmu =M_u/m_e$ and $\hdmj =
\Delta M_j/m_e = (M_i - A_i M_u)/m_e$ are the dimensionless atomic mass
unit and mass excess. We have neglected, in the original system, terms
containing the derivatives of chemical potential. The expression of the
dimensionless energy densities and pressures, $\hp_\alpha\equiv {\rm
p}_\alpha/T^4$ and $\hrho_\alpha\equiv \rho_\alpha/T^4$, contained in $G$,
were evaluated taking also into account the $\gamma$ and $e^{\pm}$
electromagnetic mass renormalization, and fitted as functions of $z$ for
their inclusion in the BBN code (see Appendix A of \cite{code}). The
previous effect, changing the $\gamma$ and $e^{\pm}$ equations of state,
modifies the $T_\nu/T$ ratio too. However, the difference between the
neutrino temperature evaluated with the correct renormalized masses and the
one obtained with approximated expressions, $m_\gamma^R \simeq 0$ and
$m_e^R \simeq \alpha T^2/ m_e$, results to be smaller than $0.01 \%$, a
correction that can be neglected at the level of precision we are
interested in. 

The initial conditions for Eq.s~\eqn{eq:bbn1} and \eqn{eq:bbn2} are given
by
\be
\hh_{in} = \frac{2 \zeta(3)}{\pi^2} \eta_{in} = \frac{11}{4}~ \frac{2
\zeta(3)}{\pi^2} \eta, 
\ee
in terms of the final baryon to photon density ratio, $\eta$, and
\bea
&&\!\!\!\!\!\!\!\!\!\!\!\! X_i(T_{in}) = \frac{g_i}{2} \left(
\zeta(3)\sqrt{\frac{8}{ \pi}} \right)^{A_i-1} \!\!\! A_i^{\frac{3}{2}} \,
\left( \frac{T_{in}}{M_N} \right)^{\frac{3}{2} (A_i-1)} \nonumber \\
&&\!\!\!\!\! \times~ \eta^{A_i-1} \, X_p^{Z_i} \, X_n^{A_i-Z_i} \,
\exp\left\{\frac{B_i}{T_{in}}\right\}, 
\eea
which represents the condition of nuclear statistical equilibrium for an
arbitrary $i$-th nuclide, with $g_i$ internal degrees of freedom, $Z_i$
and $A_i$ charge and atomic number, and $B_i$ binding energy. This
condition is satisfied with high accuracy at the initial temperature
$T_{in}=10 ~MeV$.

\subsection{Numerical Method}

The numerical problem of solving the set of equations \eqn{eq:bbn1} and
\eqn{eq:bbn2} is {\it stiff}, because the r.h.s. of \eqn{eq:bbn2} results
to be a small difference of large numbers. While in the standard code
\cite{stcode} the implicit differentiating method (backward Euler scheme)
\cite{numrec} for writing the r.h.s. of \eqn{eq:bbn2} and a Runge-Kutta
solver are used, we choose a method belonging to the class of Backward
Differentiation Formulas (BDFs) \cite{numrec}, 
implemented by a NAG routine. 

The new code includes all the 88 reactions between the 26 nuclides present
in the standard code with the same nuclear rate data, collected and updated
in \cite{network}. However, in order to reduce the computation time we used
a reduced network, made of 25 reactions between the first 9 nuclides (see
Table 1 of Ref. \cite{code}), verifying that this affects the abundances
of light nuclei for no more than $0.01~\%$. 

\section{Results and conclusions}

We report in Table \ref{tab:abund} the predictions for $Y_2 =
\frac{X_3}{X_2}$, $Y_3 = \frac{X_4}{X_2}$, $Y_4 = \frac{M_6~ X_6}{\sum_j
M_j~ X_j}$ and $Y_7 = \frac{X_8}{X_2}$, corresponding to the complete $n
\leftrightarrow p$ rates, $\omega_{Tot}$, and to the Born approximation,
$\omega_B$. This last quantity denotes the pure Born predictions for $n
\leftrightarrow p$ rates without any constant rescaling of coupling to
account for the experimental value of neutron lifetime. These values have
been obtained for $N_\nu=3$ and $\eta=5\cdot 10^{-10}$. The net effect of
the corrections is to allow a smaller number of neutrons to survive till
the onset of nucleosynthesis. This ends up in a smaller fraction of
elements fixing neutrons with respect to the pure hydrogen.

\begin{figure}
\vspace{9pt}
\epsfxsize=7cm
\epsfysize=7cm
\epsffile{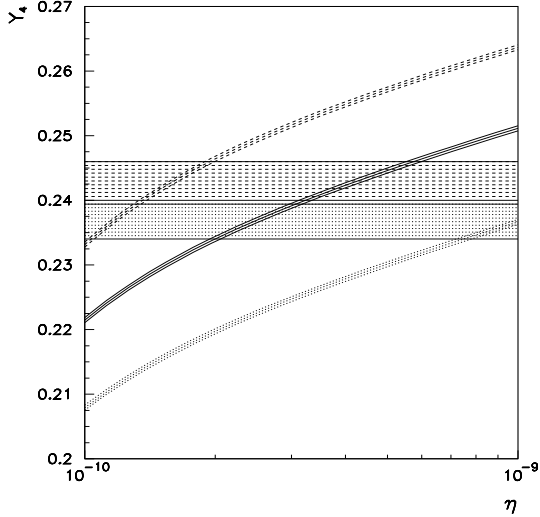} 
\caption{The $^4He$ mass fraction, $Y_4$, versus $\eta$ is shown. The
horizontal dashed and dotted bands are the experimental values.} 
\label{fig:he4}
\end{figure}

\begin{figure}
\vspace{9pt}
\epsfxsize=7cm
\epsfysize=7cm
\epsffile{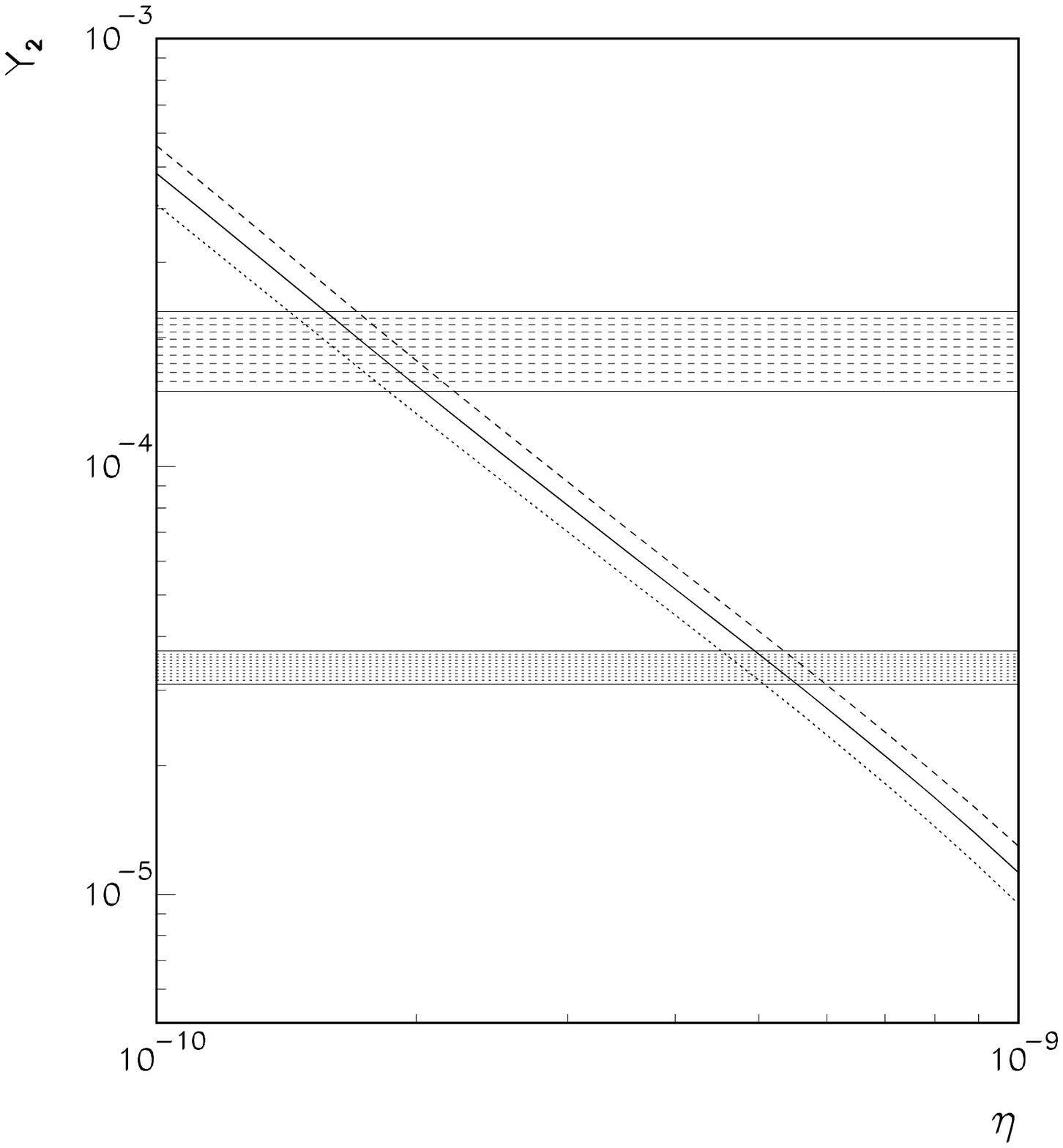} 
\caption{The quantity $Y_2$ versus $\eta$ is reported. The horizontal
dashed and dotted bands are the experimental values.} 
\label{fig:deu}
\end{figure}

\begin{figure}
\vspace{13pt}
\epsfxsize=7cm
\epsfysize=7cm
\epsffile{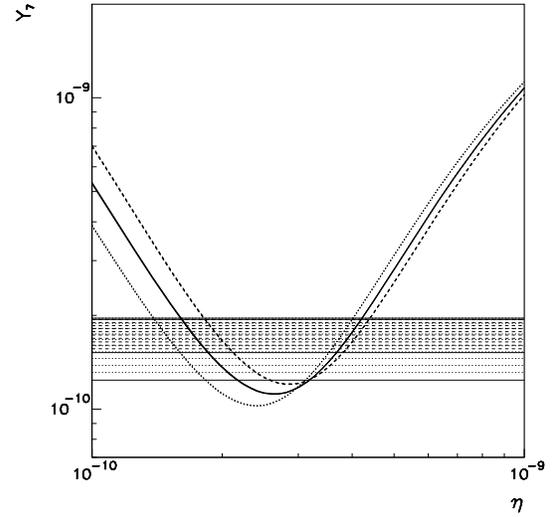} 
\caption{The quantity $Y_7$ versus $\eta$ is reported. The horizontal 
dashed and dotted bands are the experimental values.} 
\label{fig:li}
\end{figure}

In Fig. \ref{fig:he4} the predictions on $Y_4$ are shown for $N_\nu=2,3,4$
and for a $1~\sigma$ variation of $\tau_n^{ex}$. The three solid lines
are, from larger $Y_4$ to lower values, the predictions corresponding to
$N_\nu=3$ and $\tau_n^{ex}=888.6~s$, $886.7~s$, $884.8~s$, respectively.
Analogously, the dashed lines correspond to $N_\nu=4$ and the dotted ones
to $N_\nu=2$. The experimental estimates, as horizontal bands, are also
reported. In Fig.s \ref{fig:deu} and \ref{fig:li} the $D$ and $^7Li$
abundances are reported with the same notation. Note that, due to the
negligible variation of $Y_2$ and $Y_7$ on small $\tau_n$ changes, no
splitting of predictions for $1~\sigma$ variation of $\tau_n^{ex}$ is
present.

\begin{figure*}[p]
\begin{center}
\begin{tabular}{cc}
\epsfxsize=6cm
\epsfysize=6cm
\epsffile{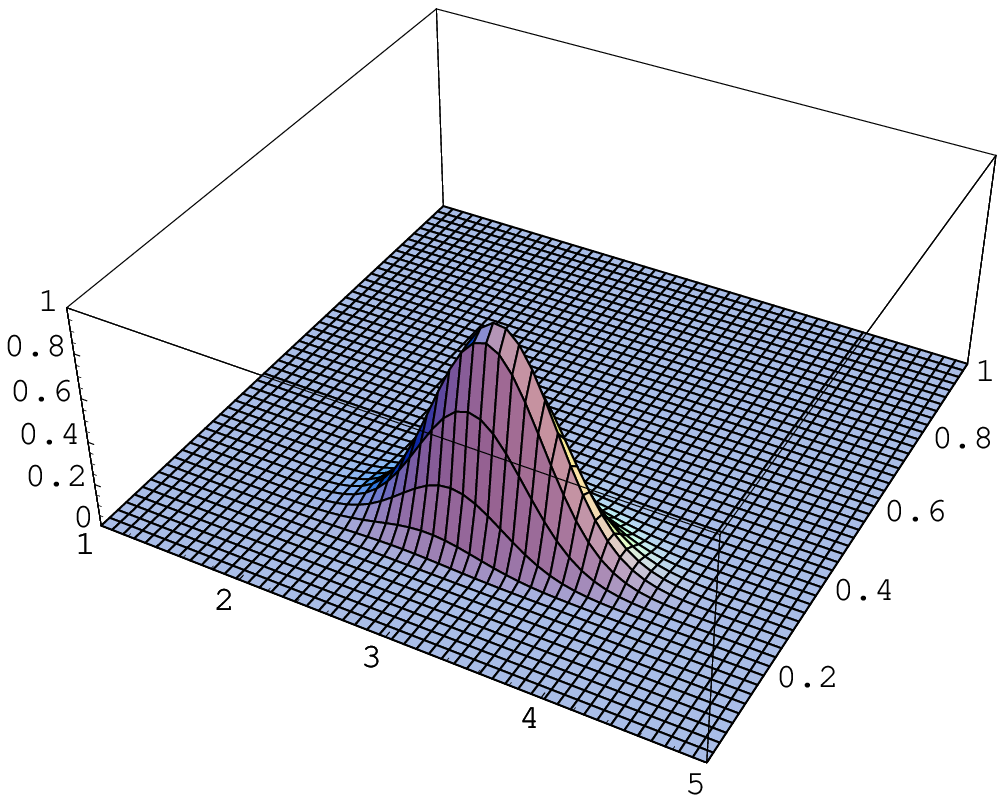} &
\epsfxsize=6cm
\epsfysize=6cm
\epsffile{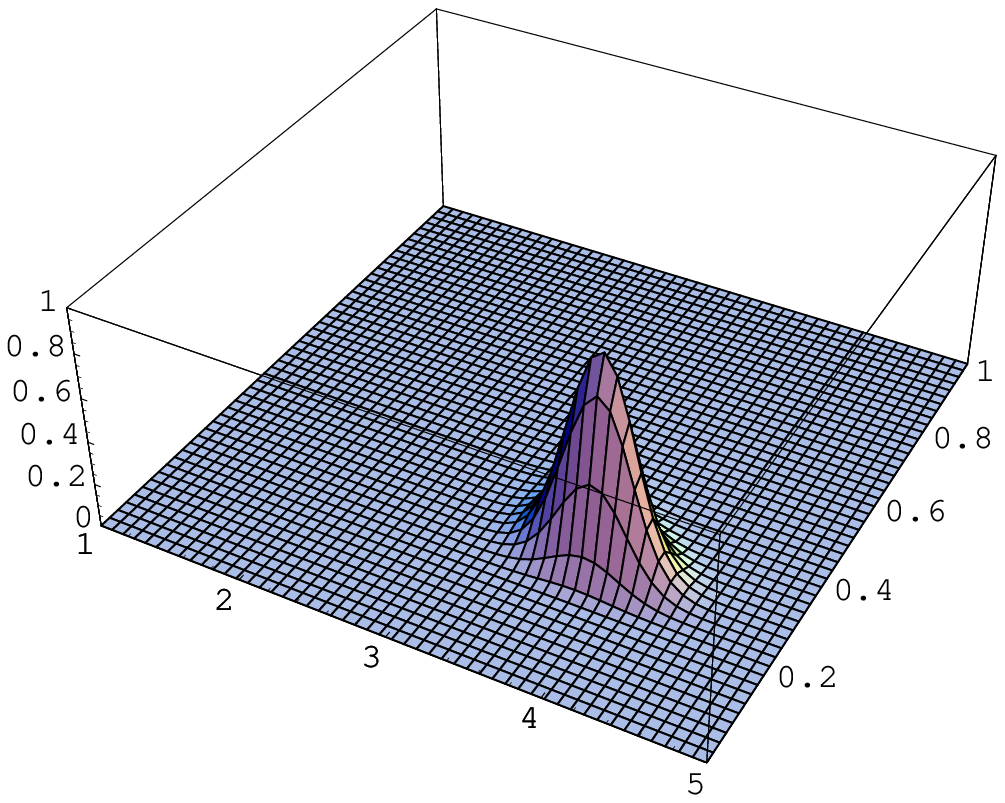} \\
\epsfxsize=6cm
\epsfysize=6cm
\epsffile{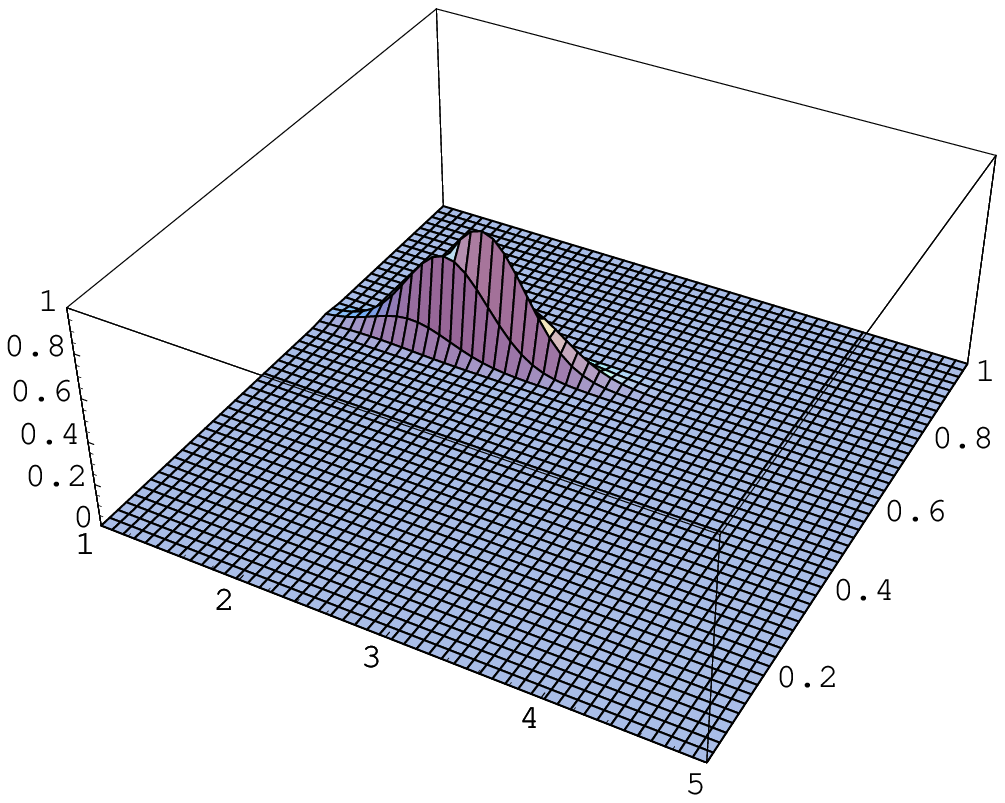} &
\epsfxsize=6cm
\epsfysize=6cm
\epsffile{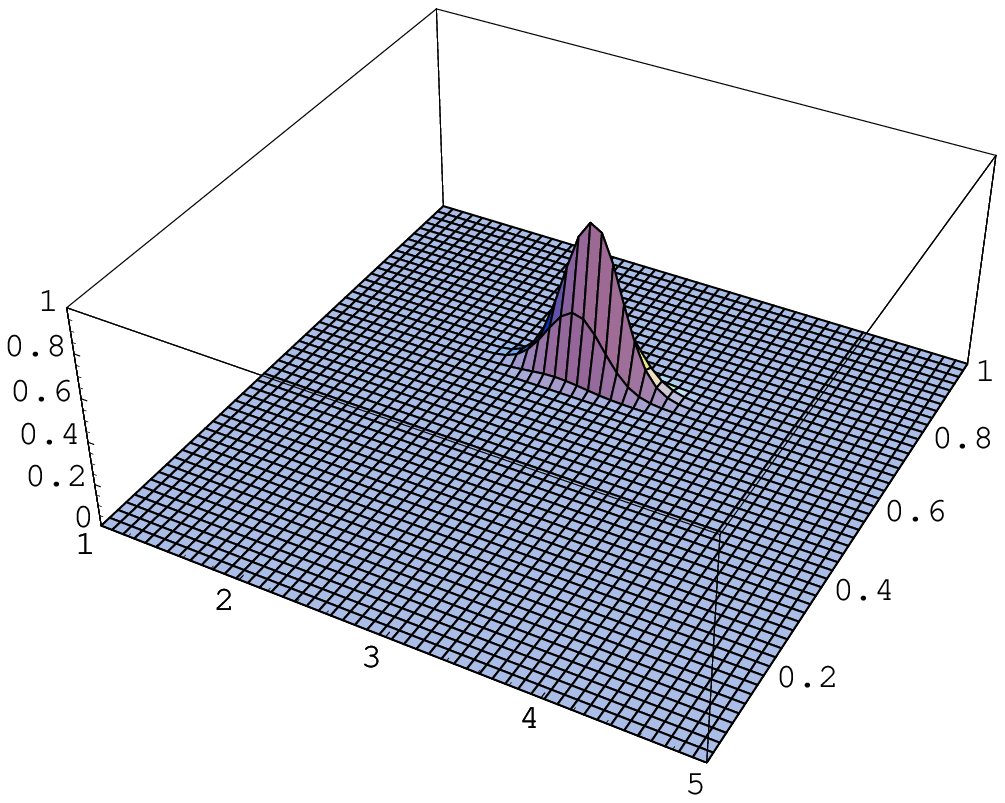}
\end{tabular}
\end{center}
\caption{The likelihood distributions for the light element yields $Y_2$,
$Y_4$, $Y_7$ are shown as functions of $N_\nu$ and $\log_{10} 10^{10}
\eta$, in arbitrary units. From left to right and from top to bottom the
following cases are considered: a) high $D$, low $^4He$; b) high $D$, high
$^4He$; c) low $D$, low $^4He$; d) low $D$, high $^4He$. The normalizations
of c) and d) are 25 and 100 times, respectively, the one of a) and b).} 
\label{fig:likeplot}
\end{figure*}

By fitting, up to one percent accuracy, $Y_2$, $Y_3$, $Y_4$ and $Y_7$ as a
function of $x = \log_{10} \left( 10^{10}~ \eta \right)$, $N_\nu$ and
$\tau_n$, the following expressions have been obtained:
\bea
&&\!\!\!\!\!\!\!\!\!\!\!\! 10^{3} {\cdot} Y_2 = \left[ \sum_{i=0}^{4} a_i
\, x^{i} + a_5 \,( N_\nu -3)\right] \,\exp\left\{ - a_6 \,x \right.
\nonumber \\ 
&& \left. + a_7 \, x^2 \right\}, \\ 
&&\!\!\!\!\!\!\!\!\!\!\!\! 10^{5}  {\cdot} Y_3 = \left[ \sum_{i=0}^{4} a_i
\, x^{i} + a_5 \,( N_\nu -3)\right] \,\exp\left\{ - a_6 \,x \right\}, \\ 
&&\!\!\!\!\!\!\!\!\!\!\!\! 10 {\cdot} Y_4 = \sum_{i=0}^{5} a_i \, x^{i} +
a_6 \,(\tau - \tau_{ex}) + a_7 \,( N_\nu -3) \nonumber \\
&& + a_8 \, x \, (\tau - \tau_{ex}) +a_9 \, x\, ( N_\nu-3), \\ 
&&\!\!\!\!\!\!\!\!\!\!\!\! 10^{9} {\cdot} Y_7 = \left[ \sum_{i=0}^{3} a_i
\, x^{i} + a_4 \,( N_\nu -3)+ a_5 \, x \,( N_\nu -3)\right] \nonumber \\
&& \times \exp\left\{ - a_6 \,x + a_7 \,x^2 + a_8 \, x^3 + a_9 \, x^4
\right\}, 
\eea
where the values of the fit coefficients are reported in Table 3 of
\cite{code}. 

In Fig. \ref{fig:likeplot} we plot the product of gaussian distribution for
$D$, $^4He$ and $^7Li$, centered around the measured values and with their
corresponding experimental errors, 
\bea
L(N_\nu,x) &=& \exp\left({(Y_2(N_\nu,x)- Y_2^{ex})^2}\over{2 \sigma^2_{2}}
\right) \nonumber \\ 
&\times& \exp\left({(Y_4(N_\nu,x)- Y_4^{ex})^2}\over{2 \sigma^2_{4}}
\right) \nonumber \\ 
&\times& \exp\left({(Y_7(N_\nu,x)- Y_7^{ex})^2}\over{2 \sigma^2_{7}}
\right). 
\eea
The experimental values used in the previous equation correspond to the
four combinations of experimental results: a) $Y_2^{(h)}$, $Y_4^{(l)}$; b)
$Y_2^{(h)}$, $Y_4^{(h)}$; c) $Y_2^{(l)}$, $Y_4^{(l)}$; d) $Y_2^{(l)}$,
$Y_4^{(h)}$. 

The figure shows that the high value of $D$ is preferred (plots a) and b)).
In both cases the distributions are centered in the range $x\in 0.2\div
0.4$, but at $N_\nu\sim 3$ for low $^4He$ and $N_\nu\sim 3.5$ for high
$^4He$. For low $D$ the compatibility with experimental data is worse (note
that in c) and d) cases the distributions have been multiplied by a factor
of 25 and 100 respectively) and centered in the range $x\in 0.6\div 0.8$,
and at $N_\nu\sim 2$ for low $^4He$ and $N_\nu\sim 3$ for high $^4He$.

\end{document}